\documentclass[12pt]{article}
\begin{document}

\begin{titlepage}

\begin{center}
\hfill hep-th/0003072\\
\hfill UM--TH/99-13

\vskip 1.5 cm
{\Large \bf On Seven-Brane and Instanton Solutions of}
\vskip .5cm
{\Large \bf Type IIB}

\vskip 1 cm

{\large Martin B. Einhorn and Leopoldo A. Pando Zayas}\\

\vskip 1cm
Randall Laboratory of Physics\\
The University of Michigan\\
Ann Arbor, Michigan 48109-1120\\
\vskip .5cm

{\tt lpandoz, meinhorn@umich.edu}

\end{center}

\vskip 0.8 cm

\begin{abstract}

It is shown that magnetic seven-branes previously considered as different objects are members of a one-parametric family of supersymmetric seven branes. We enlarge the class of seven-branes by constructing new magnetically and also electrically charged  seven branes. The solutions display a kink-like behavior. We also construct a  solution that naturally generalizes the D-instanton.

\end{abstract}

\end{titlepage}

\date{}

\def\a{\alpha}
\def\b{\beta}
\def\o{\omega}
\def\O{\Omega}
\def\d{\delta}
\def\ep{\epsilon}
\def\e{\eta}
\def\t{\tau}
\def\bt{\bar{\tau}}
\def\te{\theta}
\def\p{\phi}
\def\S{\Sigma}
\def\r{\rho}
\def\m{\mu}
\def\n{\nu}
\def\na{\nabla}
\def\g{\gamma}
\def\G{\Gamma}
\def\l{\lambda}
\def\L{\Lambda}
\def\umu{\underline \mu}
\def\unu{\underline \nu}
\def\um{\underline m}
\def\un{\underline n}
\def\pd{\partial}
\def\bpd{\bar{\partial}}
\def\beq{\begin{equation}}
\def\eeq{\end{equation}}
\def\beqa{\begin{eqnarray}}
\def\eeqa{\end{eqnarray}}
\def\dt{\tilde d}


\section{Introduction}

Solitonic solutions of supergravity have played a crucial role in achieving our current understanding of nonperturbative string theory. The identification of a subset of these solutions with the strong coupling limit of D-branes was central to the resolution of longstanding puzzles in black hole physics and large N gauge theories. Although in general the properties of this solutions are well known (see reviews \cite{ss, youm,pope,stelle}),  it is fair to say that some solutions have been  explored less than others. Such is the case of 7-branes and instantons in type IIB. Seven branes play a central role in different recent developments in nonperturbative string theory. Two examples are the instrumental role in the formulation of F-theory \cite{vafa} and in constructions of supersymmetric gauge theories \cite{gauge7}. The role of the type IIB instanton has been widely discussed in the context of the AdS/CFT correspondence.

In this paper we attempt to remedy this situation by widening the class of seven-brane solutions \footnote{Some papers considering different aspects of seven branes include \cite{ortin, lozano}.} and showing that the D-instanton of Gibbons, Green and Perry \cite{ggp} admits a very natural generalization. We also hope to clarify some misconceptions in the literature by showing how some solutions that have been regarded as independent of each other are actually particular cases of a one-parametric family of supersymmetric seven-brane solutions. 

In section 2 we study various seven-brane constructions. After reviewing the most general solution of the seven-brane ansatz and showing how it reduces to previously known solutions we study seven branes from a holomorphic point of view, this is a distinguishing feature of seven-branes that have been widely exploited in nonperturbative constructions. We show that holomorphic seven branes are very similar to the nonlinear $O(3)$ model used to describe isotropic ferromagnets. In particular we show explicitly that by imposing very reasonable asymptotic conditions on the solution, one is led to a homotopical classification for the charge of the seven-brane, {which is not possible for D-branes in general.}  In the two last subsections of section 2 we construct new seven-brane solutions, we consider both electric and magnetic

seven-branes. The gauge field of the electrically charged seven-brane displays a kink-like behavior. In section 3 we discuss the supersymmetry of seven branes and show that these equations admit the one-parametric family of solutions constructed at the beginning of section 2. In section 4 we construct generalizations of the D-instanton of \cite{ggp}.

\section{On seven-branes}
\subsection{The various D7-branes}
In this subsection we review the known seven-brane solutions and clarify a common misconception about the D7-brane. We argue  that three seven-brane constructions found in the literature are basically the same object. The seven-brane constructed implicitly in \cite{ss,pope} along the lines of the traditional p-brane solution, the GGP seven brane \cite{ggp} and a {\it circular} seven-brane constructed in \cite{circular} are all, to an extent to be specified shortly, particular cases of a unique construction. Let us start by reviewing the general p-brane construction in 10-dimensions. One starts with the following Einstein-frame action \cite{ss,pope}
\beq
S_{II,p}={1\over 16 \pi G}\int d^{10}x\sqrt{-g}\left(R-{1\over 2}(\pd\p)^2-{1\over 2 (p+2)!}e^{{3-p\over 2}\p}F_{p+2}^2\right).
\label{p-action}
\eeq
The equations of motion (EOM) for $p=-1$ are:
\beqa
R_{MN}&=&{1\over 2}(\pd_M\p\pd_N\p+e^{2\p}F_MF_N), \nonumber \\
{1\over \sqrt{-g}}\pd_M(\sqrt{-g}g^{MN}\pd_N\p)&=&e^{2\p}g^{MN}F_MF_N,\nonumber\\
\pd_M(\sqrt{-g}g^{MN}e^{2\p}F_N)&=&0.
\label{eqm}
\eeqa
Here, $F_M=\partial_M{\rm a},$ where ${\rm a}$ is the zero-form, R-R potential.  Unlike the quantum theory, the classical theory is invariant under a rescaling of the action, and, as the Lagrangian has no mass parameter, this feature is critical for understanding some of the properties of classical solitons, such as D-branes, in the absence of other scales.

One expects the D7-brane to be a singular, magnetic solution of $S_{II,-1}$ and the corresponding ansatz for the metric and field strength is \cite{ss,youm,pope}
\beqa
ds^2&=& e^{2A(r)}\e_{\m\n}dx^\m dx^\n+ e^{2B(r)}\d_{mn}dy^mdy^n, \nonumber \\
F_m&=&\l\ep_{mn}{y^n\over r^2}\qquad {\rm or}\qquad  a=\lambda\theta,
\label{metgauge}
\eeqa
with $\m,\n=0,\ldots, 7$ ,\, $m,n=1,2$; $\ep_{12}=1$; $r=\sqrt{y_1^2 + y_2^2}$; $\theta$ is the angular coordinate in the transverse space. This ansatz for the metric is dictated by preserving $P_8\times SO(2)$, that is, Poincare invariance in the longitudinal directions and $SO(2)$ in the perpendicular ones. One very useful relation, which is true for all Dp-branes and is required by supersymmetry is $A~d+B~\tilde{d}=0$, where $d=p+1$ and $\tilde{d}=10-p-3$  \cite{ss} (see appendix for the general definition). Applied to seven-branes $(\tilde{d}=10-8-2=0),$ one has $A=0$. The Ricci tensor in this case becomes $R_{mn}=-\d_{mn}\pd^2B$ (see appendix). Einstein's equations then imply
\beqa
\p'&=-&{\l e^{\p}\over r}, \nonumber \\
B''+{B'\over r}&=&-{1\over 2}(\p')^2.
\eeqa
The first of these equations is characteristic of magnetic p-brane solutions and says that $\exp{-\p}$ is a harmonic function in the transverse coordinates. The general solution to the second equation can be written as
\beq
B=B_0+B_1\r+{1\over 2}\ln({1+\l\r}),
\eeq
where we have defined $\r\equiv\ln(r/r_0)$, a notation that we shall find useful throughout the rest of this paper.  By a rescaling of $y^m$, we may choose $B_0=0$, whence the entire background takes the form
\beqa
ds^2&=&\e_{\m\n}dx^\m dx^\n+e^{2(B_1+1)\r}(1+\l\r)(d\r^2+d\te^2), \nonumber\\
e^{-\p}&=&1+\l\r ,\nonumber \\
F_m&=&\l\ep_{mn}{y^n\over r^2}.
\label{d7}
\eeqa

For sufficiently small $r,$ viz., $r<r_1\equiv r_0\exp{(-1/\l)}$, the quantity  $1+\l\r$ can become negative, and the solution breaks down, since of course $\exp(-\p)$ must be nonnegative.  This also shows up as a singularity in the metric and, more importantly, as a naked singularity in the curvature.
\beq
R={\l^2\over{r^{2(B_1+1)}(1+\l\r)^3}}
\eeq
In this respect, the 7-brane differs from lower p-branes, whose singularities do not occur at finite $r.$  This disturbing feature of the solution can and should be interpreted as the appearance of new physics in a natural way: The SUGRA Lagrangian, being nonrenormalizable, must be understood as the first term of an effective field theory involving an expansion in powers of the curvature so, as the curvature becomes large, the terms ignored become as important as the leading term.  The value of $\r_1$ is arbitrary because the Lagrangian is scale-invariant.  However, by reference to other forms of matter, it is natural to expect such a breakdown on a scale $r\sim G^{1/8}.$  Similarly, from the superstring  point of view, one expects corrections at $r\approx \sqrt{\a'}.$  Moreover, since $\exp(\p)$ is the (local) string coupling constant, this becomes large as $r$ approaches $r_1,$ so that one cannot trust the classical approximation anyway.   Therefore, even though in principle $r_0$ is simply an integration constant, we assume that it is sufficiently small so that our solution makes sense physically at distances $r>>r_1$.

For $B_1=0,$ one has the typical {harmonic function} representation for p-branes \cite{ss,youm,pope}. For $B_1=-1$, one recovers the circular seven-brane of \cite{circular}.\footnote{To obtain precisely eq.~(6.14) in \cite{circular} it is necessary to make the following change: $\ln r\to r$ and to identify $\l=\tilde{m}$.} It is not possible to establish a one-to-one relation with the solution of \cite{ggp}. One can however show that the asymptotic behavior of the solution of \cite{ggp} can be obtained from eq.~(\ref{d7})\footnote{The limit we consider here has been previously discussed in the literature (see, for example,  \cite{dabholkar}). This limit is known as the weak coupling limit of the GGP solution.} It is very natural to concentrate on the large $r$ asymptotic behavior because most of the constraints that string theory imposes through the boundary state formalism on supergravity solutions depend only on the asymptotic behavior. The notational correspondence between the asymptotic behavior of the dilaton and the axion field given in \cite{ggp} and the fields of eq.~(\ref{d7}) are:
\beq
\l={1\over 2\pi}, \qquad b={1\over{r_0}}.
\eeq
The asymptotic value of the dilaton $b$ appears in eq.~(28) of \cite{ggp}.

To establish the relation between the metrics we now turn to eq.~(27) in \cite{ggp} (see next subsection)
\beq
\pd\bpd\ln\O={1\over 2}\pd\bpd \ln \t_2 .
\eeq
where, in our notation, $\ln\O=B$ and $\t_2=\exp{(-\p)}.$ The general solution to this equation is ${\rm{e}}^{2B}=\t_2\exp(F(z)+\bar F(\bar{z}))$. where $F$ is an arbitrary holomorphic function. Rotational invariance in the transverse dimensions leads to the choice (\ref{d7}). Further specification within this class amounts to different choices of the parameter $B_1$, as shown before. The conditions imposed in \cite{ggp} were that the metric must be  real, modular invariant and nondegenerate (for an extensive discussion see~\cite{yau}). This brings us to $\exp(2B)=(1+\l\ln({r\over {r_0}}))\e^2\bar{\e}^2$, where $\e$ is the Dedekind eta function, which is perfectly compatible with (\ref{d7}) for asymptotically large $r$. To guarantee nondegeneracy of the metric  in the presence of multiple seven-branes, other powers of $r$ must be included but this is still compatible with (\ref{d7}).

\subsection{Comments on holomorphic seven-branes}

In the case $p=-1$ the action (\ref{p-action}) can be rewritten as
\beq
S_{II, -1}={1\over 16 \pi G}\int d^{10}x\sqrt{-g}\left(R-{1\over 2\t_2^2}\pd_\m \t\pd^\m\bt \right),
\label{complex}
\eeq
where $\t\equiv\t_1+i\t_2=a+ie^{-\p}$. One interesting property of the equation of motion following from this action is that, for the type of metric we are considering here ($A=0$,) the dilaton and axion equations are independent of the metric.  This decoupling from the metric signals a complex structure that has been exploited in various contexts starting from~\cite{yau}. The equation of motion for $\t$ is
\beq
\pd\bpd \t+{2\pd\t\bpd\t \over \bt-\t}=0,
\eeq
where $z\equiv y^1+iy^2$ and $\partial\equiv\partial_z.$  One can easily see that any holomorphic or antiholomorphic function $\t$ is a solution to this equation. This particular class of solutions is required for supersymmetric solutions (see below). Einstein's equations take the form\beq
\pd\bpd B={\pd\t\bpd\bt+ \bpd\t\pd\bt \over 2(\t-\bt)^2}={1\over 2}\pd\bpd \ln\t_2,
\eeq
where the last equation is true only for holomorphic or antiholomorphic $\t$. At this stage one can conclude that the seven-brane-like solutions to the EOM obtained from (\ref{complex}) are characterized by an arbitrary holomorphic function $F(z)$.
\beq
e^{(2B)}=\t_2e^{2(F+\bar F)}.
\label{hol}
\eeq
This situation is familiar, it is typical of systems with two spatial dimensions, such is the case of the nonlinear $O(3)$ model~\cite{raja} that describes isotropic ferromagnets. In our problem, however, we have to go a step further because not any holomorphic function is acceptable as opposed to the case of the $O(3)$ model.


The domain in which $\t$ is defined presumably carries some information about the symmetries of the theory. The idea follows from the fact that in supergravity $\t$ belongs to the upper half plane $\bf H$ and the symmetry of the theory is $SL(2,R)$. This  symmetry is expected to be broken at the string theory level due to charge quantization to a discrete subgroup, $SL(2,Z),$ which is presumed to be a discrete gauge symmetry.  In that case, $\t$ must be restricted to the fundamental domain; this was the path taken in~\cite{ggp} which is asymptotically compatible with eq.~(\ref{d7}) and raises the natural question of how much supergravity solutions know about the full string theory. We will explore the restrictions that must be applied to $\t$ such that a well-defined magnetic charge exists at the end of this subsection.

If one insists on a metric preserving $P_8\times SO(2)$, which is the usual ansatz for the p-brane metric, then using Einstein eq.~(\ref{hol}), one has that $\t_2=\t_2(r^2)$. The condition that the sum $F+\bar F$ be a function of $r^2=z\bar z$ poses a functional equation stating that the sum of $F$ and $\bar F$ must be expressed as a product of their arguments. This equation is solved for $F(z)=B_1\ln z$. Using holomorphicity of $\t: \bpd \t=0$ we get: $i\dot a/\bar z+a'z+i\t_2'z=0$, where the dot means derivative with respect to $\te$ and the prime  with respect to $r^2$. For purely magnetic solutions, $(a'=0)$, one has that $\t_2'=-\dot a/(z\bar z)$. Since $\t_2'$ is a function of only $r^2=z\bar z$ and $\dot a$ is a function of only  $\te=-(i/2)\, \ln z/\bar z$, the only solution is $\dot a=-\l$ which implies $\t_2=1+\l\ln r$. This is essentially the solution presented in eq.~(\ref{d7}).

The converse is however not true.  For any magnetically charged holomorphic seven-brane: $i\dot a/(2\bar z)+i\t_2'z-\dot\t_2/(2\bar z)=0$. Since $a$ and $\t_2$ are  real, the only solution to this equation is $\dot \t_2=0$ ({\it i.e.,} $\t_2$ is independent of the angle.)  For the same reasons, $\dot a=const=\l$ and $\t_2=\exp({-\p})=1+\l\r.$  This, however, is not enough to fix the form of $F,$ and the solution will in general depend on an arbitrary holomorphic function.

The analogy with the $O(3)$ model can be pursued a step further to establish that the charge is equal to the degree of the map $\t\to z$. For $\exp(-2\pi i\t)=bz^n+c$, one has that for asymptotically large $r$, $a=-n\te/(2\pi)$ which implies $q=-n$, the same relation one has for the $O(3)$ model. This means that the charge equals the homotopy class of the map $\t\to z$ and provides the homotopic sector classification known for solitons and instantons in gauge theories. This is a distinctive feature of seven-branes, the gauge fields in D-branes are $U(1)$  gauge fields and therefore the only chance to get some nontrivial homotopy is for two-dimensional transverse space due to $\pi_k(U(1))=0$ for $k>1$. It is interesting to note that, as in solitons and instantons in gauge theory, the long distance behavior determines the topological class of the solution on very general grounds and that one need not know the details of the solution for all values of the radius.

\subsection{Electric kink-like seven brane}

In this section we construct an electric\footnote{Here the term electric is used exclusively in the sense that $a=a(r)$ and, unlike the previous case, the conserved charge $\int_{S^1}e^{2\p}(\pd_r a)$ is nonzero.} seven-brane with a metric somewhat resembling the one of the circular seven-brane constructed in~\cite{circular}.

We know from general considerations \cite{polchinski, ggp} that the charge dual to the D7-brane charge (called ``magnetic" here) is the D(-1)-brane charge, i.e., the Type~IIB-instanton charge that couples locally to the zero-form $a$ field.  The instanton, as a classical solution of the EOM, only exists in Euclidean space-time, although of course it represents a quantum tunneling amplitude in a Hilbert space built up in a space-time with Lorentz signature. The charge that the solution we present here carries plays a prominent role in the D-instanton solution of \cite{ggp}. One interpretation of this charge will be evident at the end of this section.

We will look for solutions of the truncated IIB supergravity Lagrangian discussed in the previous sections. The main difference will be that, in the ansatz for the metric, we will not assume $A=0$ as we did previously, so, as mentioned earlier, such a solution will not be supersymmetric and not BPS.\footnote{Hence, for strong coupling, we can expect quantum corrections to be large.}  The reason for starting with this general ansatz is that for $A=0,$ there are no electric solutions. From the Einstein equations eq.~(\ref{eqm}) and the general form of the Ricci tensor (see appendix) one has, in the transverse directions, $R_{mn}=-\d_{mn}\pd^2B$,  but the right hand side of Einstein's equations is $y_my_n/r^2((\p')^2+e^{2\p}(a')^2)$ which has a different tensor structure and is signaling  that the equation can not be nontrivially satisfied. We will, therefore, consider the following ansatz
\beq
ds^2= e^{2A(r)}\e_{\m\n}dx^\m dx^\n+ e^{2B(r)}\d_{mn}dy^mdy^n.
\eeq
The EOM are given by eq.~(\ref{eqm}). The Einstein equation for $g^{\m\n}$ implies that
\beqa
\partial^2e^{8A}&=&0, ~{\rm or~more~explicitly,} \nonumber \\
A''+\frac{1}{r}A'+8(A')^2&=&0.
\eeqa
The solution of this harmonic equation is
\beq
A-A_0=\frac{1}{8}\ln\r,
\label{a}
\eeq
where, as before, $\r=\ln(r/r_0).$  The Einstein equation in the direction transverse to the brane has two different tensor structures: $\d_{mn}$ and $y^my^n/r^2$. The $\d_{mn}$ equation gives
\beq
B''+\frac{1}{r}B'+8A'B'+\frac{8}{r}A'=0.
\eeq
Solving this equation is straightforward, giving
\beq
B-B_0=-\r+B_1\ln\r ,
\eeq
here $B_1$ is an arbitrary constant. Using the axion equation, one can rewrite the remaining Einstein equation (for $y^my^n/r^2$) in terms of the dilaton only
\beq
\p''+\frac{1+\r^{-1}}{r}  \p'+(\p')^2=(4B_1+\frac{7}{4})\frac{\r^{-2}}{r^2}.
\eeq
This equation may be linearized by replacing $\p$ by $\exp(\p)$ and further simplified by defining $t\equiv \ln\r$, yielding
\beq
{{d^2}\over{dt^2}}e^\p=\o^2e^\p,
\eeq
Remarkably, this is invariant under translation of $\ln (r/r_0)$ (or rescaling of $\r$,) but this is clearly not a symmetry of the other equations.
The solution of this equation may be expressed in a variety of ways, e.g.,
\beq
e^\p=\cosh(\o(t-t_0)) = {1\over2}\left[ \left({\r\over\r_0}\right)^\o + \left({\r_0\over\r}\right)^\o\right]
\eeq
where $\o=\sqrt{4B_1+7/4}$, $\r_0$ is a constant and $t=\ln\rho$.  Up to this point, we actually have not found it necessary to have $\o^2>0$ but, from this solution, we see that this is required in order to have $\exp(\p)\ge 0$  for all $t.$\footnote{This corresponds to the restriction $B_1\ge -7/16.$}

The general solution for the metric and the axion is, up to a global $SL(2,R)$ transformation,
\beqa
ds^2&=&(\r )^{1/4}\e_{\m\n}dx^\m dx^\n+
r_0^2\r^{2B_1}(d\r^2+d\te^2),\nonumber \\
 a&=&\tanh\left(\o\ln({\r\over{\r_0}}) \right),
\eeqa
where $\r_0$ is a constant.  This has a form very similar to the circular seven-brane of~\cite{circular}.

It is easy to see that $|\t|=1,$ so that the solution corresponds to a classical solution in the $\t$-plane that starts at $\t=-1$ and traverses a semicircle until arriving at $\t=+1.$  

We would like to stress at this point, although it will be completely clear in what follows, that the solution presented here, which we called electric seven brane because the charge that it carries is a Noether charge, i.e. it is conserved as a consequence of the equations of motion, is not a supersymmetric solution. In the absence of supersymmetry or a  topologically conserved charge, it is not at all clear that this solution is stable. One of the properties of the charge of the electric seven brane presented here is that it is invariant under the $SL(2,R)$ transformation acting on the dilaton and action fields, viz., $\tau \to (a\tau+ b)/(c\tau +d)$. This means that the electric charge can be viewed as an $SL(2,R)$ charge. A similar situation takes place for the instanton soltuion of \cite{ggp}. In this context the electric seven brane and the D-Instanton are two $SL(2,R)$-charged objects.  

\subsection{Magnetic seven-branes}

In this subsection we will construct general magnetic seven-branes. The ansatz for the metric and the field strength is exactly as at the beginning of this section, eq.~(\ref{metgauge}). The Einstein equation of motion in the longitudinal directions to the brane  is as in the previous subsection and therefore the solution for $A$ is as for the electric 7-brane eq.~(\ref{a}), $A-A_0=(1/8)\ln\ln r/r_0$. The ``off-diagonal" Einstein equation 
\beq
(\p')^2-{e^{2\p}\l^2 \over r^2}=-16(A''-{A'\over r})-16(A')^2+32A'B'.
\label{offdiag}
\eeq
The two ``diagonal" equations are identical
\beq
{1\over 2}(\p')^2+8A''+B''+{B'\over r}+8(A')^2-8A'B'=0.
\label{diag}
\eeq
The axion equation, as is always the case for magnetic solutions, is identically satisfied. The dilaton equation becomes
\beq
\p''+{\p'\over r}+8A'\p'={\l^2e^{2\p}\over r^2}.
\eeq
A convenient way to solve these equations is, once again, to change variables to $\r=\ln r/r_0$. In this new coordinate the dilaton equation becomes
\beq
\p''+{\p'\over \r}=\l^2e^{2\p},
\eeq
where the prime now denotes the derivative with respect to $\r$ . Using eq.~\ref{offdiag}, equation~\ref{diag} takes the following simple form in terms of $\r$
\beq
B''+{B'\over \r}+{1\over \r}+{\l^2e^{2\p}\over 2}=0.
\eeq
The previous two equations can be combined using the identity $(\p'\r)'/\r=\p''+\p'/\r$ into
\beq
B'+{\p'\over 2}={t_0\over \r}-1,
\eeq
where $t_0$ is an integration constant.  These results can be plugged into eq.~\ref{offdiag} giving
\beq
\p''-{\p'\over \r}-(\p')^2=-{(16t_0+7)\over 4\r^2}.
\eeq
A linearization procedure, similar to the one discussed in the electric case, can be applied to this equation with the following modification 
\beq
{{d^2}\over{dt^2}}e^{-\p-t}=\L^2e^{-\p-t},
\eeq
The complete solution takes the following form
\beqa
ds^2&=&\r^{1/4}\e_{\m\n}dx^\m dx^\n+r_0^2\r^{2t_0+1}\sinh\Big(\L\ln ({\r\over{\r_0}})\Big)\left(d\r^2+d\te^2\right),\nonumber \\
e^{-\p}&=&e^{-\p_o} \r\sinh\Big(\L\ln({\r\over{\r_0}}) \Big), \nonumber \\
a&=&-\l \te,
\eeqa
with the following algebraic constraint, $\l e^{\p_0}=\L,\,\, \L^2=4t_0+11/4$. 
We could just set $\phi_0=0$ so that $\L$ becomes $\l$.
We shall now show that there is a limit in which this solution becomes the solution presented in eq.~(\ref{d7}). In order to do so let us consider the following substitution  $8\ln(r_0)=-1/C_O$. Using this substitution and renaming some of the integration constants we can rewrite
\beqa
A-A_0&=&{1\over 8}\ln(1+C_0\ln r),\nonumber \\
B-B_0&=&-\ln r +(t_0+1/2)\ln(1+C_0\ln r)-{1\over 2}\ln\sinh(x_0+\L\ln(1+C_0\ln r)),\nonumber \\
\phi-\phi_0&=&-\ln (1+C_o\ln r)-\ln\sinh(x_0+\L\ln(1+C_0\ln r)).
\eeqa
The limit to the ``extremal" case eq.~(\ref{d7}) is: $C_0, \,x_0 \to 0$ and $\L,\, e^{2B_0},\,\,e^{-\phi_0}\to \infty$ in such a  way that $C_0\L^2,\, C_0\L/x_0, \, e^{2B_0}x_0$ and $\exp{-\phi_0}x_0$ are constant. This can be easily arranged by sending  $\ep$ to zero in the following relations
\beq
C_0=\ep^2, \quad \L^2={B_1 +1\over \ep^2}, \quad x_0= \sqrt{B_1+1}{\a\over \m}\ep, \quad e^{2B_0}=e^{-\phi_0}={\m\over \sqrt{B_1+1}}{1\over \ep}.
\eeq
This way we recover precisely the solution of eq.~(\ref{d7}).

\section{Comments on supersymmetric seven-branes}

In this section we will analyze the condition under which a seven-brane solution admits a supersymmetric generalization. The supersymmetry transformations of the corresponding fermionic fields in type IIB supergravity can be presented in an $SU(1,1) $ \cite{schwarz,west1} (see also the recent review \cite{west2}) or $SL(2,R)$ invariant formulation \cite{west3,iib}.

The dilaton and axion fields of type IIB theory parametrize
the upper half of the complex plane using the traditional parametrization $\t=a+ie^{-\p}$. Following  \cite{schwarz,west1,west2,west3,iib}, one can introduce the zweibein $V^\a_{\pm}$,
\beq
  V = \left( \matrix{
      V^1_- & V^1_+ \cr V^2_- & V^2_+} \right),
\label{matV}
\eeq
and define the local $U(1)$ action as
\beq
 V \rightarrow V \left( \matrix{ e^{-i\Sigma} & 0 \cr
         0 & e^{i\Sigma} } \right),
\label{uone}
\eeq
where $\Sigma$ is a $U(1)$ phase.
It is convenient to parametrize the matrix $V$ following~\cite{iib}
\beq
 V = \frac{1}{\sqrt{-2i\tau_2}}
 \left( \matrix{ \bar{\tau} e^{-i\g} & \tau e^{i\g} \cr
       e^{-i\g} & e^{i\g} } \right).
\label{param}
\eeq
One can fix the local $U(1)$ gauge symmetry by setting
the scalar field $\g$ to be a function of $\tau$: $\g=\g(\t)$

To write the type IIB supergravity EOM, it is convenient to introduce
two $SL(2,R)$ singlet currents \cite{schwarz,west1,west2,west3,iib},
\beqa
  P_M & = & - \epsilon_{\alpha\beta} V_+^\alpha \partial_M
 V_+^\beta
= \frac{i}{2} \frac{\partial_M \tau}{\tau_2} e^{2i\g}, \nonumber \\
Q_M & = &
-i \epsilon_{\alpha\beta} V_-^\alpha \partial_M V_+^\beta
 = \partial_M \g - \frac{1}{2} \frac{\partial_M \tau_1}{\tau_2}.
\label{currents}
\eeqa
Under the $U(1)$ gauge symmetry (\ref{uone}), they transform
as
\beqa
  P_M & \rightarrow & P_M e^{2i\Sigma}, \nonumber \\
  Q_M & \rightarrow & Q_M + \partial_M \Sigma.
\eeqa
These transformations show that $Q_M$ is a composite gauge potential and that the $U(1)$ charge of $P_M$ equals 2.

The equations that result from sending all the fermions to zero as well as all bosonic fields except the relevant ones 
are \cite{schwarz,west1,west2,west3}
\beqa
 R_{MN} &=& P_M P^*_N + P^*_M P_N, \nonumber \\
 D^M P_M &=&  (\nabla_M - 2i Q_M) P_M = 0.
\eeqa
Substituting (\ref{currents}) into these, the EOM
can be expressed in terms of $\phi$ and $a$ as
\beqa
&& R_{MN} = \frac{1}{2} (\partial_M \phi \partial_N \phi
+ e^{2\phi} \partial_M a \partial_N a), \nonumber \\
&&  \Delta a + 2 \partial^M \phi \pd_M a = 0, \nonumber \\
&&  \Delta \phi - e^{2\phi} (\partial a)^2  = 0.
\eeqa
Note that $\g$ does not appear in the EOM. In fact it appears as a multiplier in the dilaton and the axion equations but this does not affect the EOM. It should be taken as a bonus that these equations can be derived from a Lagrangian because the original definition of IIB  is solely based on supersymmetry and the EOM. The Lagrangian in question is precisely the one we have been studying given by eq.~(\ref{p-action}).

The supersymmetry transformations of the dilatino $\l$ and
the gravitino $\psi_M$ are given by
\beqa
  \d \l &=& i P_M \g^M \ep^*, \nonumber \\
  \d \psi_M &=& \left( \nabla_M - \frac{i}{2} Q_M
\right) \ep .
\label{susy}
\eeqa
From the supersymmetric variation of the dilatino it is easy to see that a supersymmetric solution must have holomorphic or antiholomorphic $\t$. Namely,
\beq
\d\l=-{e^{2i\g-B}\over 2\t_2}(\G^1\pd_1\t+\G^2\pd_2\t)\ep^*=-{e^{2i\g-B}\over 2\t_2}\G^1(\pd_1\t+\G^1\G^2\pd_2\t)\ep^*.
\eeq
Taking $\G^1\G^2\ep^*=i\ep^*$ one obtains that in order for the variation of the dilatino to be zero, $(\pd_1+i\pd_2)\t=0$, which means that $\t$ is a holomorphic function. The same analysis could be carried out for antiholomorphic functions.

For the gravitino one has that the variation in the longitudinal directions is
\beq
\d\psi_\m=\e_{\m\umu}e^{A-B}\G^{\umu}\G^m\pd_m A\ep,
\eeq
here we have assumed that all fields depend only on the transverse directions. One could apply the same type of manipulation we applied to the dilatino transformation and conclude that a solution to this equation is provided by any antiholomorphic function $A$, but $A$ is a metric entry and must, therefore, be real. The only choice we have is $A=constant$ which, by means of a trivial coordinate redefinition, is equivalent to $A=0$. So far, everything is compatible with the type of seven-branes considered in subsection 2.1. The gravitino transformation in the transverse directions is
\beq
\d\psi_m=\left(\pd_m+{1\over 2}\pd_nB\G^m\G^n-{1\over 2}\pd_mB-{i\over 2}\pd_m \g+{i\over 4\t_2}\pd_m\t_1\right)\ep.
\eeq
Substituting $B=B_0+B_1\ln r+{1\over 2}\ln \t_2$, taking into account that $\t$ is a holomorphic function and considering a more general spinor $\ep=e^{if/2}\ep_0$ where $\G^1\G^2\ep_0^*=i\ep^*_0$, we obtain that $f-\g+iB_1\ln r$ must be a holomorphic function or in other words, $\g-f=-B_1\te$ which can always be satisfied for any given $\g=\g(\t)$. In the case of the D7-brane of~\cite{ss,pope} one has $B_1=0$ which is a solution for $\g=0$ (the preferred gauge fixing condition) and $f=0$. The solutions of~\cite{ggp} and~\cite{circular} also enter in this scheme. To make contact with~\cite{ggp} we need to drop the condition that $B=B(r)$ and assume only that $B$ must be real. In this case the solution is specified up to a holomorphic function eq.~(\ref{hol}). The variation of the gravitino in the directions perpendicular to the seven-brane demand that $f-\g+i(F+\bar F)$ be a holomorphic function. In~\cite{ggp} $\g(\t)=-\mbox{ Im}\ln(\t+i)$ up to an additive  constant, this combines with the fact that $F=F(\t)$ is dictated by modular invariance, to provide a supersymmetric solution for $f=0$.

\section{Instantonic solutions to IIB}
In this section we will consider the following Euclidean action
\beq
S={1\over 16 \pi G}\int d^{10}x\sqrt{g}\left(R-{1\over 2}(\pd\p)^2+ {1\over 2 }e^{2\p}(\pd a)^2\right).
\eeq
and its EOM
\beqa
R_{MN}&=&{1\over 2}(\pd_M\p\pd_N\p-e^{2\p}F_MF_N), \nonumber \\
{1\over \sqrt{g}}\pd_M(\sqrt{g}g^{MN}\pd_N\p)&=&-e^{2\p}g^{MN}F_MF_N, \nonumber\\
\pd_M(\sqrt{g}g^{MN}e^{2\p}F_M)&=&0.
\eeqa
One way to generalize the solution of~\cite{ggp} is to consider the following ansatz for the metric
\beq
ds^2=e^{2B(r)}\d_{mn}dy^mdy^n.
\label{imetric}
\eeq
As familiar by now, the Einstein equations have two tensor structures. The $\d_{mn}$ part is
\beq
B''+\frac{17}{r}B'+8(B')^2=0,
\eeq
which becomes a linear differential equation after substituting $B'=1/u$. The general solution is
\beq
B-B_0=\frac{1}{8}\ln(1+\frac{Q_0}{r^{16}}).
\eeq
Note that for $Q_0=0$ and fixing the constant $B_0=0$ we recover the flat metric of the solution obtained in~\cite{ggp}.

The dilaton and axion equation take the following form
\beqa
&&\p''+{9\over r}\p' + e^{2\p} (a')^2+ 8  B'\p'=0, \nonumber \\
&& a'' +{9\over r}a'+ 2a'\p'+8B'a'=0.
\eeqa
Using the axion equation one obtains that $y^my^n/r^2$ Einstein equation can be rewritten as
\beq
\p''+(\p)^2+\p'({9\over r} +8B')={288 \over Q_0}r^{16}(B')^2.
\eeq
In terms of a new function $D$ such that $\p'=(D-4B'r^9)/r^9$  and also changing to coordinate $t: {dt\over dr}=1/r^9$ the dilaton equation becomes
\beq
\dot{D}+D^2=-{512 Q_0 \over (1+64Q_0t^2)^2}.
\label{dil}
\eeq
The solutions of this equation take different forms depending on the sign of  $Q_0$. For negative  $Q_0$, substituting $Q_0\to -Q_0^2$ in all formulas we obtain the following solution:
\beqa
ds^2&=&(1-{Q_0^2\over r^{16}})(dr^2+r^2d\O_9^2),\nonumber \\
e^\p&=&e^{\bar \p_0}\sinh\left(x_0+{3\over 2}\ln\frac{1+Q_0/ r^8} {1-Q_0/ r^8}\right),\nonumber \\
a-a_0&=&-e^{-\bar \p_0}\coth\left(x_0+{3\over 2}\ln\frac{1+Q_0/ r^8}{1-Q_0/ r^8}\right).
\label{n}
\eeqa
The limit in which this solution coincides with~\cite{ggp} is the following: $Q_0,\,x_0\,\to 0$ and $\exp{\p_0}\to \infty$ with the following relations going to constants: $Q_0/x_0=C/3,\,\, e^{\bar \p_0}x_0=e^{ \p_0}$. (Here $C$ and $\p_0$ are the parameters of the  solution in~\cite{ggp}.) This solution has a singularity at $r_0^8=Q_0$. It is a nontrivial task to classify this singularity being a singularity in Euclidean space. Let us simply note that it is a curvature singularity since for this metric the scalar curvature equals
$$
R={288Q^2_0\over r^{18}} (1-{Q_0^2\over r^{16}})^{-9/4}.
$$

There is also a solution for positive values of the parameter $Q_0$, for this case we will consider $Q_0\to Q_0^2$. The solution we obtained is
\beqa
ds^2&=&(1+{Q_0^2\over r^{16}})(dr^2+r^2d\O_9^2),\nonumber \\
e^\p&=&e^{\p_0}(1+{C\over r^8}-{3Q^2_0\over r^{16}}-{CQ^2_0\over 3r^{24}})(1+{Q^2_0\over r^{16}})^{-3/2},\nonumber \\
a&=&e^{-\p_0}( 1-\frac{ \sqrt{C^2+9Q_0^2}-C }{3r^8}\frac{ 3r^{16}-Q_0^2}{r^{16}-3Q_0^2} )/( 1+\frac{C}{3r^8}\frac{3r^{16}-Q^2_0}{r^{16}-3Q^2_0}).
\eeqa
To recover the instanton of~\cite{ggp} from this solution one needs simply send $Q_0\to 0$.

The Einstein equations of $S_{II,-1}$ (both for Minkowskian and Euclidean versions) imply that the action is zero. This naive approach would render an instanton with zero action. First of all one would  turn to the surface term. The surface term originates from allowing variations of $\d g^{MN}$ which vanish at the surface but whose normal derivatives do not. The boundary term is $\int (K-K^0)\sqrt{h}d^9y$. For the dilaton metric we consider here one obtains a contribution proportional to $O(1/r^8)$ which therefore vanishes for asymptotically large distances. A way out of this situation which was successfully used in~\cite{ggp} consists in realizing that the same solution can be considered as a solution of the supergravity but instead of considering an axion field one needs to consider the  $C_8$ formulation with field strength $F_{9}$ dual to the field strength of the axion $da$. In this picture Einstein equation no longer implies that the solution must have zero action.   There is a more rigorous argument as to why the Euclidean version for the axion field has to include the surface term, it is based on an analysis of the dual formulation \cite{ggp,gg,beh}. A subtlety here arises because of the need to perform Poincare duality to objects that live in spaces of different signatures. Modulo these subtleties our result for the value of the action is exactly that of~\cite{ggp}. We checked that our solution becomes that of~\cite{ggp} in certain limit specified by the charges. It is straightforward to notice that this limit corresponds precisely to the large $r$ limit of the solution in the positive $Q_0$ case and for negative $Q_0$ we need to assume that $x_0=0$.

\section{Conclusions}


In this paper we have constructed new electric and magnetic seven-branes. We have shown that the EOM for the seven-brane ansatz admit a one-parametric family of solutions some particular cases of which were known in the literature as separate objects. We have also shown explicitly that the supersymmetry equations are loose enough to admit a solution for such one-parametric family. The electric and magnetic seven-branes constructed here have a kink-like behavior similar to some solitons in gauge theories.

We have constructed instantonic solutions for IIB that generalize in a very clear way the previously known D-instanton solution of~\cite{ggp}. We have not, however, analyzed the supersymmetric structure of the instanton.
\vfill\eject

\begin{center}
{\large \bf Acknowledgments}
\end{center}
We wish to thank M. Duff for pointing out a solution we overlooked at the early stage. We would also like to thank  J.T. Liu and J.X. Lu for very interesting discussions and several clarifying remarks.  We would also like to acknowledge useful correspondence from Y. Lozano and M. Gutperle. One of us (L. PZ) would like to thank the Office of the Provost at the University of Michigan for support.  Both of us wish to acknowledge the support of the High Energy Physics Division of the Department of Energy.

\vspace{.8cm}

{\Large \bf Appendix}

\vspace{.8cm}

The calculations for the Ricci tensor for the p-brane metric have been presented in a number of papers \cite{ss, stelle, pope}. We present such results here for the sake of completeness. Let us consider a metric in $D$ spacetime dimensions, given by the following ansatz
\beq
ds^2 = e^{2A}\, dx^\mu dx^\nu \eta_{\mu\nu} +
       e^{2B}\, dy^m dy^n \delta_{mn}\ ,
\eeq
where $x^{\mu}$ $(\mu = 0, \ldots, d-1)$ are the coordinates of the
$(d-1)$-brane world volume, and $y^m$ are the coordinates of the
$(D-d)$-dimensional transverse space.   Generally the functions $A$ and $B$ depend
only on $r=\sqrt{y^my^m}$. Here we will consider an arbitrary dependence in the transverse dimensions. The Ricci tensor obtained from this metric is
\beqa
R_{\mu\nu} &=& -\eta_{\mu\nu} e^{2(A-B)} \d^{mn}\left ( \pd_m\pd_n A+d\,\pd_mA\pd_nA+\dt\,\pd_mA\pd_nB\right),\nonumber\\
R_{mn} &=& -\d_{mn}\left(\d^{pq}\pd_p\pd_qB+d\,\d^{pq}\pd_pA\pd_qB+\dt\,\d^{pq}\pd_pB\pd_qB\right)\nonumber \\
&&-\dt\,\pd_m\pd_nB-d\,\pd_m\pd_nA-d\,\pd_mA\pd_nA+\dt\,\pd_mB\pd_nB\nonumber \\
&&+d\,(\pd_mA\pd_nB+\pd_nA\pd_mB),
\eeqa
where $\dt= D -d - 2$.   A convenient choice of vielbein basis for the metric
is $e^{\underline \mu} = e^{A} d x^\mu$ and $
e^{\underline m} = e^{B} dy^m$, where underlined indices denote tangent
space components.  The corresponding spin connection, which is needed for the supersymmetry transformation of the gravitino,  is
\beqa
\omega^{{\underline\mu}{\underline n}} &=& e^{-B}\pd_n A\, e^{\underline \mu}
\ ,\qquad \omega^{{\underline \mu}{\underline \nu}} = 0\ ,\nonumber\\
\omega^{{\underline m}{\underline n}} &=& e^{-B} \pd_n B\, e^{\underline m}
-e^{-B} \pd_m B\, e^{\underline n}.
\eeqa

\end{document}